\documentclass[12pt]{article}

\begin{document} 
\def\ba{\begin{eqnarray}}

\def\ea{\end{eqnarray}}

\def\lb{\label}

\def\nn{\nonumber \\}

\def\bi{\bibitem}

\def\g{\gamma}

\def\d{\delta}

\def\D{\Delta}

\def\ee{\epsilon}

\def\o{\omega}

\def\a{\alpha}

\def\m{m_\perp}
\def\pu{p_{\bar{u}}}
\def\pd{p_{\bar{d}}}
\def\q{\bar{q}}
\def\r{\rho}
\def\w{\bar{w}}
\def\rr{\rightarrow}
\def\uu{\bar{u}}
\def\dd{\bar{d}}
\def\c{\cos}

\def\f{\phi}

\def\e{\eta}

\def\t{\theta}

\def\n{\bar{n}}

\def\F{\bar{F}}

\def\ff{\bar{f}}

\def\pp{p_\parallel}
\def\pt{p_\perp}

\def\l{\lambda}

\def\ss{\vec{s}}

\title{ Balance functions revisited}

\author{A.Bialas  \\ H.Niewodniczanski 
Institute of Nuclear Physics\\
Polish Academy of Sciences\thanks{Address: Radzikowskiego 152, Krakow,
Poland}\\and\\
M.Smoluchowski Institute of Physics \\Jagellonian
University\thanks{Address: Reymonta 4, 30-059 Krakow, Poland;
e-mail:bialas@th.if.uj.edu.pl;}}

\maketitle

keywords: balance functions, short-range correlations, glue clusters 

\begin{abstract}

The idea of glue clusters, i.e. short-range correlations in the
quark-gluon plasma close to freeze-out, is used to estimate the width of
balance functions in momentum space. A good agreement is found with the
recent measurements of STAR collaboration for central $Au-Au$
collisions.

\end{abstract}

\vspace{0.3cm}

\section {Introduction}

It was observed by Bass, Danielewicz and Pratt \cite {bdp} that
measurements of balance functions in nucleus-nucleus collisions provide
a "clock" which allows to determine when the charges observed in the
final state are created. The measurements performed by STAR
\cite{starold,starnew,st} and NA49 \cite{na1,na2} collaborations proved
(or at least strongly suggested) that the charges appear at the late
stage of the production process, close to the freeze-out
\cite{prattetal}, thus implying that the quark-gluon plasma phase is
dominated by gluons.

In addition, these experiments showed that the measured widths of the
balance functions in central $AA$ collisions are substantially smaller
than those seen in $pp$ and in peripheral $AA$ collisions. The natural
conclusion from this observation is that the presence of the quark-gluon
plasma phase induces additional correlations in the system. The nature
of these correlations is of course a matter of debate. They were
extensively discussed \cite{prattetal} in the framework of the thermal
blast wave model \cite{blast}. It was argued that, when the quark-gluon
plasma is sufficiently cool, pairs of positive and negative charges are
created, with thermal momentum distribution, in certain restricted
domains in space regions (as implied by the HBT measurements
\cite{hbt}). The resulting momentum separation is measured through the
balance function. Since the creation of charges happens close to the end
of the evolution, the separation of charges induced by expansion of the
system is not effective and thus a small width of the balance function
is maintained.

In \cite{ab}, on the other hand, the additional correlations were
interpreted as evidence for clustering in the quark-gluon plasma. It was
shown that production of uncorrelated isotropic clusters of gluons
decaying into $q\q$ pairs, when supplemented by their coalescence into
hadrons \cite{cm}, explains -in a natural way- the correct width of the
balance function in pseudorapidity.

In the coalescence model \cite {cm} it is assumed that hadrons are
created by coalescence of "constituent" quarks and antiquarks into
$q\q$, $qqq$ and $\q\q\q$ colour singlets. If there are no correlations
in the system before coalescence, then the only correlations which can
appear between hadrons can be those induced by resonances (which also
result from coalescence). This picture describes reasonably well the
hadron-hadron data \cite{foa}. The additional short-range correlations,
needed to account for the experimental experimental measurements of STAR
\cite{starold,starnew,st} and NA49 \cite{na1,na2} collaborations, may
only appear if the constituent quarks and antiquarks are correlated
between themselves before the coalescence process takes place. The
postulated glue clusters supply these additional correlations.

Since pseudorapidity is determined solely by the particle production
angle, distributions in pseudorapidity are only marginally sensitive to
particle momenta. It is therefore interesting to verify if the
hypothesis of glue clusters can explain also the recently measured
\cite{starnew} balance functions in momentum space. This is the subject
of the present investigation. Our main conclusion is that, indeed, the
presence of glue clusters at the last stages of the evolution of the
quark-gluon plasma can account for the observed width of balance
funtions in $q_{side}$, $q_{long}$ and $q_{out}$. This result requires
that the average momentum $<q>$ carried by a quark and antiquark in the
decay of a glue cluster is located around $\sim 120$ MeV.

In the next section we derive the momentum dependence of the balance
function. Comparison with data is presented in Section 3. Our
conclusions and comments are listed in the last section. The Appendix
summarizes our treatment of acceptance corrections.

\section {Balance function from  glue clusters}

Consider  a two-body decay of a  glue cluster
 into a $q\bar{q}$ pair. The distribution of the decay products is
\ba
\frac{dn}{dp_1dp_2}= v_c(p_1-p_2)\delta(p_1+p_2-P) \lb{qmm}
\ea
where $P$ is the momentum of the cluster, $p_1$ and $p_2$ are momenta of
the decay products and the function $v_c$ is responsible for the details
of the decay ($dp$ stands for three-dimensional relativistic
phase-space, $dp=d^3p/E=d^2p_\perp dy$).

 Consequently,  the distribution of quarks and antiquarks 
arising from two clusters is
\ba
\frac{dN(p_u,p_{\uu},p_d,p_{\dd})}{dp_ud\pu dp_d d\pd}=
\int dP_UdP_D \r_c(P_U)\r_c(P_D)\nn
\d(p_u+\pu-P_U)v_c(p_u-\pu)
\d(p_d+\pd-P_D)v_c(p_d-\pd)
\ea
where $\r_c$ is the distribution of clusters in momentum  space.
The subscripts $U,D$ refer to the flavour of quarks to which the cluster
decays, and $u,d,\bar{u},\bar{d}$ to flavour of the quarks and
antiquarks themselves. 

For the  balance function we obtain\footnote{Since we are only
interested in the width of the balance function, henceforth 
 we shall ignore all normalizing factors.}
\ba
B(p_+,p_-)\sim\int dp_ud\pu dp_d d\pd \d(p_u+\pd-p_+)\d(p_d+\pu-p_-)\nn
\Phi(p_u-\pd)\Phi(p_d-\pu)
\frac{dN(p_u,p_{\uu},p_d,p_{\dd})}{dp_ud\pu dp_d d\pd}
\ea
where $\Phi(x)$ describes the coalescence process.

Introducing
\ba
p_+=p_u+\pd\;;\;\;\; p_-=\pd+p_u\;;\;\;\;\ \d_+=p_u-p_{\bar{d}}\;
 ;\;\;\;\d_-=p_d-\pu
\ea
we have
\ba
B(p_+,p_-)\sim  \int d\d_+d\d_- \r_c(p_u+\pu)\r_c(p_d+\pd)
\Phi(\d_+)\Phi(\d_-)\nn
v_c(p_u-\pu)v_c(p_d-\pd)\d(p_u+\pd-p_+)\d(p_d+\pu-p_-)
\ea
and thus
\ba
B(\d)\sim  \int d\d^+d\d^- \Phi[\d^++\d^-)/2]\Phi[\d^+-\d^-)/2]\nn
\int dp \r_c[(p+\d^-)/2]\r_c[(p-\d^-)/2]
v_c[(\d+\d^+)/2]v_c[(\d-\d^+)/2]   \lb{bfinal}
\ea
where  
\ba
p=p_++p_-;\;\;\; \d=p_+-p_- ;\;\;\;\d^\pm=\d_+\pm\d_-\;.
\ea
Eq. (\ref{bfinal})  represents our final result. Its consequences for the
widths of the balance functions in various components of momentum are
discussed in the next section.

\section {The widths of balance functions}

To simplify the discussion and reduce the number of parameters,
we assume in this section that the functions
$v_c(x)$ and $\r_c(x)$ are Gaussians:
\ba
v_c(x)\sim e^{-x^2/v^2};\;\;\;\r_c(
x)=e^{-x_\parallel^2/r_\parallel^2-x_\perp^2/r^2}.
\ea
Then the dependence on $\d^+$ and $\d^-$ in (\ref{bfinal}) factorizes
out and we have
\ba
B(\d)\sim e^{-\d^2/2v^2}\Omega(\d);\;\;\;
\Omega(\d)=\int dp e^{-p_\parallel^2/2r^2_{\parallel}}
e^{-p_\perp^2/2r^2}  \lb{bcor}
\ea

If there are no acceptance restrictions, the integral extends over full
phase-space, $\Omega(\d)$ is a constant, and the balance function is
isotropic: its width does not depend on the chosen
direction\footnote{This was already observed in \cite{prattetal}.} and
is entirely determined by the parameter $v^2$. In the STAR experiment
\cite{starnew}, however, the acceptance corrections are important. They
were estimated and are summarized in the Appendix. 

Eq.(\ref{bcor}) contains three  parameters: $v$, $r$
and $r_\parallel$. This number can be still reduced, 
using the available information on the distribution of transverse and
longitudinal momenta. Indeed, the single particle transverse 
 momentum distribution can be expressed as
\ba
D(p_{+\perp})=e^{-2p_{+\perp}^2/(r^2+v^2)}d^2p_{+\perp}\rr
<p_{+\perp}^2>=(r^2+v^2)/2
\ea
with an analogous formula  for $p_\parallel$. Consequently
we have
\ba
r^2=2<p_{\perp}^2>-v^2;\;\;\; r_\parallel^2=2<p_{\parallel}^2>-v^2
\ea
and we are left with one parameter, $v^2$ which should be adjusted to
data.

One sees from these formulae that $r_\parallel$ is very large and
therefore its exact value is not important for estimate of $\d_{long}$.
Therefore, to determine $v$, we used the value $ < \d_{long}> = 190 $
MeV, measured in the central $Au-Au$ collisions \cite{starnew}. Using
Eq. (\ref{distlong}) from the Appendix,
 one obtains $v=276$ MeV. With this value and 
$<p_{\perp}> = 400$ MeV \cite{phe}, Eqs. (\ref{distside})
 and (\ref{distout}) give
\ba
<\d_{side}>=284\ MeV;\;\;\;\;  <\d_{out})>=126\ MeV
\ea
 This should be compared with 
\ba
<\d_{side}>=280\pm 10\ MeV;\;\;\;\;
<\d_{out})>=0.110\pm 10\ MeV,
\ea
 measured in \cite{starnew}. Given the crudeness of
our (Gaussian) approximation, the agreement is more than satisfactory.
The width uncorrected for acceptance is $<\d> = 220$ MeV.

From the obtained value $v=276$ MeV and using (\ref{qmm}) one can
evaluate the average value $<q>$ of the momentum carried by the quark
and antiquark in the decay of the cluster. One obtains
$<q>=v\sqrt{\pi}/4\approx 122$ MeV and $\sqrt{<q^2>}=v/2\approx 138$
MeV.

\section {Conclusions and comments}

In conclusion, we verified that the hypothesis of glue clusters, i.e.
positive short-range correlations in the quark-gluon plasma \cite{ab}
can account for the small widths of the balance functions in momentum
space, observed recently by the STAR collaboration \cite{starnew} for
central $Au-Au$ collisions. In particular, it was shown that when the
parameters of the model are determined from the observed $<\d_{long}>$,
one obtains reasonable  values of $<\d_{side}>$ and  of
$<\d_{out}>$. This result confirms the existence of correlations in the
plasma. It also shows that they can be effectively studied by the method
of balance functions \cite{bdp}.

Several comments are in order. 

(i) It should be emphasized that the observed small width of balance
functions in momentum space simply {\it implies} the existence of
additional short range correlations between particles produced in
central nucleus-nucleus collisions (as compared to pp collisions). The
nature and origin of these correlations remains a subject of debate but
their very existence is beyond doubt. For example, the explanation of
the balance functions in the blast wave model \cite{blast}, presented in
\cite{prattetal}, also exploits the correlations in the plasma (in the
form of "domains" from which the balancing charges are emitted).

(ii) The existence of short-range correlations in the quark-gluon plasma
close to the phase transition  should not be surprizing.
Indeed, the lattice calculations indicate strong deviations from the
Stefan-Boltzman limit even at temperatures largely exceeding $T_c$
\cite{latt}. Thus the presence of quasiparticles is likely. The
nature of these quasiparticles is an open question. Our results
suggest that they  may take the form of gluonic clusters.

(iii) Although our semi-analytic estimates are admittedly rather crude,
it is interesting to observe that the average momentum characterizing
the decay of a glue cluster $(120$ MeV) seems close to the break-up
temperature in the blast wave model, as discussed in \cite{prattetal}.
This may suggest that the approach of \cite{prattetal} and our
interpretation may not impossible to reconcile. Precise data on balance
functions for $KK$ and $K\pi$ pairs may throw some light on this
problem.

(iv) In our argument we have used the coalescence model \cite{cm}.
Although the model is rather successful in explaning, e.g., the particle
content \cite{cm1} and $v_2$ scaling \cite{volo}, its basic idea is
often questioned since it is difficult to reconcile with the picture of
pion as the Goldstone boson of chiral symmetry breaking \cite{blaizot}.
A possible way out is to admit that the constituent quark mass depends
on the temperature of the system (or rather on its distance from the
temperature the chiral transition). If true, then also the mass of the
glue clusters will depend on the temperature. Although our results do
not depend on the constituent quark mass, it may be interesting to
investigate this problem in more detail.

(v) It was noticed in \cite{prattetal} that corrections due to
production of hadronic resonances may change the results by 10-20 \%.
This is not dramatic for our semi-quantitative approach but must be
eventually improved if serious comparison with data is attempted. 

It is well known that resonance production is essential in description
of the short range correlations in pp collisions \cite{foa} and that it
allows to explain the single particle spectra in nucleus-nucleus
collisions (see e.g. \cite {bf}). Furthermore, it has been shown
\cite{bww} that thermal, uncorrelated production of resonances cannot
reproduce the small width of balance functions in central
nucleus-nucleus collisions. It should be noticed, however, that if
hadronic resonances are formed by coalescence of $q$ and $\q$ which are
decay products of glue clusters, they are correlated. This -in turn-
should produce additional correlations between their decay products and
thus reduce the width of the balance functions. It would be very
interesting to investigate such a possibility in detail. This, however,
requires much more sophisticated analysis than the one presented here.

(vi) The present work represents only a first order approximation to the
problem. We only considered the simplest case when the hadron pairs are
created by coalescence of quarks and antiquarks which are decay products
of just two clusters. The contributions from more than two clusters
should certainly be included in a more precise analysis, although it
seems likely that they are suppressed because (i) the probability to
have more clusters very close in space (as required by the coalescence
mechanism) is small and (ii) matching the colour of quarks from two
different clusters reduces this contribution even further. At present we
can only mention that (a) for three clusters the width of balance
function depends on details of the coalescence process and therefore its
estimate involves more parameters than hitherto considered, and (b) if
clusters are uncorrelated, contributions from 4 clusters cancel in the
balance function, see e.g. \cite{ab}.

Let us also add that the Gaussian approximation is certainly rather
crude and our treatment of the acceptance corrections is at best
approximate. This can of course be improved, although we feel that it
would be not justified at the present level of understanding.

In summary, the present investigation shows that the hypothesis of glue
clusters in quark-gluon plasma can account for the recently measured
widths of the balance functions in central $Au-Au$ collisions. This may
have important consequences for the phenomenology of the quark-gluon
plasma. It remains an open question if this result is confirmed when
data in larger acceptance regions are available.

\section {Appendix: Acceptance corrections}

We first consider transverse directions, i.e. two-dimensional vectors
$p_+$ and $p_-$. The distribution is  given by (\ref{bcor})
with the acceptance limits 
\ba
D^2\leq p_{+x}^2+p_{+y}^2 \leq \D^2;\;\;\;
D^2\leq p_{-x}^2+p_{-y}^2 \leq \D^2  \lb{acc}
\ea
where $D$= 200 MeV and $\D$= 600 MeV.

We are interested in the distribution of $\d_{out}$ and
$\d_{side}$, defined with respect to total momentum of the pair.
Since the system is invariant with respect to common rotation of all
vectors, one can select the $y$ axis as "out" and the $x$ axis as
"side". Thus we are looking for  the configuration  in which 
\ba
p_{x+}+p_{x-}=0\;\rr\; \d_{side}=\d_x=p_{x+}-p_{x-}=2p_{x+};
\;\;\;p_{\pm y}\geq 0.
\ea

Consider first $\d_{out}=\d_y$. This is the most complicated case
because in this direction there is a significant effect of the flow
which strongly distorts the spectrum and must be taken into account.
Flow implies that the cluster moves (on average) with the velocity $V$
of the fluid. Therefore the laboratory system in which it is observed
moves with the velocity $-V$ and the distribution of the decay products
is  
\ba
e^{-(\delta_{out}')^2/2 v^2}e^{-(p_{out}')^2/2 r^2};\;\;
\delta_{out}'=p_+'-p_-';\;\;p'_{out}=p_+'+p_-'
\ea
where
\ba
p_\pm'=\g[p_\pm -VE_\pm];\;\;\; E_\pm =\sqrt{p_\pm^2+x^2+m_\pi^2}
\ea
 Consequently, the observed 
distribution of the balance function is (we abbreviate $p_{+x}\equiv x$)
\ba
P(\d_{out})=\int_0^\D dx e^{-2x^2/v^2}
\int_{p_{min}}^{p_{max}}
e^{-\delta_{out}'^2/2 v^2}e^{-p_{out}'^2/2 r^2} \lb{distout}
\ea
where $p_{max}=2\D(x)-\d_{out}$, $p_{min}=2D(x)-\d_{out}$ for
$\d_{out}\leq 2D(x)$, $p_{min}=0$ for $\d_{out}\geq 2D(x)$ and
\ba
\D(x)=\sqrt{\D^2-x^2};\;\;\;D(x)=\sqrt{D^2-x^2}
\ea
Numerical evaluation of (\ref{distout}) shows that, with  $V=0.6$, 
the  effect of flow reduces significantly the width.

For $\d_{side}=\d_x$ the effect of flow is largely cancelled and we
neglect it. Consequently, the distribution is
\ba
P(\d_{side}) =e^{-\d_{side}^2/2v^2}\int_{D(x)}^{\D(x)} dp_{+y}dp_{-y}
e^{-(p_{+y}+p_{-y})^2/2r^2}e^{-(p_{+y}-p_{-y})^2/2v^2}
\ea
where $x=\d_{side}/2$.  One gaussian integration can be expressed in
terms of  error functions and one obtains 
\ba
P(\d_{side})=e^{-\d_{side}^2/2v^2}
\int_{D(x)}^{\D(x)}dy_+e^{-2y_+^2/(r^2+v^2)}\nn
\left\{erf(s|B_+|)+erf(s|B_-|)-
\left[\frac{A_+}{|A_+|}erf(s|A_+|)-\frac{A_-}{|A_-|}erf(s|A_-|)\right]
\right\}  \lb{distside}
\ea
with
\ba
A_\pm=\pm D(x)+w;\;\;\;B_\pm=\pm
\D(x)+w;\;\;\;w=y_+\frac{r^2-v^2}{r^2+v^2};\;\;s^2=\frac{r^2+v^2}{2r^2v^2}.
\ea
Note that we always have $B_+\geq 0$ and $B_-\leq 0$.

Finally, let us consider $\d_{long}$. In this case there is no lower
limit on particle momenta. 
Consequently, the distribution becomes
\ba
P(\d_{long}) \sim e^{-\d_{long}^2/2v^2}\int_0^{2\D-\d_{long}}
dp_{out}e^{-p_{out}^2/2r_\parallel^2}=\nn=
e^{-\d_{long}^2/2v^2}erf\left(\frac{2\D-\d_{long}}{r\sqrt{2}}
\right) \lb{distlong}
\ea

\section{Acknowledgements} Thanks are due to J.-P. Blaizot,
K.Fialkowski, M. Praszalowicz and K.Redlich for interesting comments.
This investigation was supported in part by the grant N N202 125437 of
the Polish Ministry of Science and Higher Education (2009-2012).

\end {document}